\documentclass[aps,preprint]{revtex4}%
\usepackage{amsfonts}
\usepackage{amsmath}
\usepackage{amssymb}
\usepackage{graphicx}%
\providecommand{\U}[1]{\protect\rule{.1in}{.1in}}

\begin{document}
\preprint{hep-th/0706.1603}
\title[Short title for running header]{Solutions of Bethe-Salpeter equation in QED$_{3}$}
\author{Yuichi Hoshino}
\affiliation{Kushiro National College of Technology,Otanohike Nishi 2-32-1,Kushiro
City,Hokkaido 084,Japan}
\pacs{PACS number}

\begin{abstract}
To understand the mechanism of the fermion pair and fermion-antifermion pair
condensation,the solutions of Bethe-Salpeter equation in QED$_{3}$ is
examined.In the ladder appoximation our solution for the axial-scalar is
consistent with Ward-Takahashi-identity for the axial vector currents.Since
the massless scalar-vector sector is described by a coupled integral
equation,it is difficult to solve explicitly.We approximate the equation for
large and small momentum region separately and convert them into differential
equations in position space.These equation can be solved easily.Boundary
condition at the origin leads the eigenvalue for dimensionless coupling
constant $\lambda=e^{2}/m$.There exists solutions for massless scalar-vector
fermion-antifermion (fa) system with discrete spectrum.In our approximation
massless-scalar-vector ff systemes does not seem to exist.

\end{abstract}
\volumeyear{year}
\volumenumber{number}
\issuenumber{number}
\eid{identifier}
\date[Date text]{date}
\received[Received text]{date}

\revised[Revised text]{date}

\accepted[Accepted text]{date}

\published[Published text]{date}

\startpage{101}
\endpage{102}
\maketitle
\tableofcontents

\section{Introduction}

In the theory of superconductivity and superfluidity BCS model is very
familiar and useful to analyse their properties[1].If we assume the
$\delta^{(3)}(x-y)$ function type intearction between fermions it is not
difficult to solve the gap equation and its dependence of the coupling
constant.However we have not been known the reason why the electron pair form
a bound state.On the other hand there exisists solutions of Bethe-Salpeter
equation in the ladder approximation for fermion-fermion pair with discrete
spectrum[2].Recently a bound state for quasi partices(exiton) are considered
in terms of approximate Bethe-Salpeter(BS) equation in QED$_{3}$ for phase
flucuating d-wave superconductor near the node to measure the resonant spin
response[3].The Schr\"{o}dinger type equation for particle-hole boundstate
with potential $1/r^{2}$ was derived and the eigenvalue condition emerged.Its
solution indicates the exsitence of strong coupling phase but seems to be not
normalizable.Therefore it is interesting to apply relativistic BS equation for
fermion pair in the same approximation in[2].We think that it is important to
solve the equation for the massless boundstate which signals the instability
of the vacuum under condensation of these bosons in field thoretical
model[4].In this work first we examine the existence of solutions for the
massless boundstates which are normalizable.Since scalar-vector sector is
written by coupled integral equation,it is difficult to solve the equation
explicitly.Therefore we approximate the equation for the large and small
momentum region separately.These equations are easily solved in position space
and we have a correct short and long-distance behaviour.For fa system the
baundary condition at the origin leads dimensionless coupling constant
$\sqrt{2}e^{2}/4\pi m=$ integer where $m$ a fermion mass.In section II we
introduce spinor-spinor BS equation and show their solutions in QED and
QED$_{3}$[2].In section III we compare the solutions of Dyson-Schwinger
equation for the fermion propagator with axialscalar solutions of BS
equation.In our approximation they obey the same equation and
Ward-Takahashi-identity for axial currents is satisfied.Section IV is devoted
for summary.

\section{Spinor-Spinor BS equation}

\subsection{Massless boson in QED}

BS amplitude in four-dimension is defined[6,7,8]%
\begin{equation}
\chi(x_{1}x_{2}\text{:}B)\equiv\left\langle 0|T(\psi(x_{1})\overline{\psi
}(x_{2}))|B\right\rangle ,
\end{equation}
for the fermion-antifermion bound state $|B\rangle=|fa,P_{\mu}\rangle$ with
total four momentum $P_{\mu}$ or as%
\[
\chi(x_{1}x_{2}\text{:}B)\equiv\left\langle 0|T(\psi(x_{1})\overline{\psi}%
^{C}(x_{2})|B\right\rangle
\]
for the fermion-fermion bound state $|B\rangle=|ff,P_{\mu}\rangle,$where
$\psi^{C}$ stands for the charge conjugated field of $\psi$.Homogeneous
Bethe-Salpeter equation for fermions in the ladder approximation is written in
the following form%
\begin{equation}
\chi(x_{1}x_{2}\text{:}B)=-e^{2}\int d^{4}x_{3}d^{4}x_{4}S_{F}(x_{1}%
-x_{3})\chi(x_{3}x_{4}\text{:}B)S_{F}(x_{2}-x_{4})\gamma_{\mu}D_{F}^{\mu\nu
}(x_{3}-x_{4})\gamma_{\nu}.
\end{equation}
We can also write the BS equation in a differential form by applying Dirac
operator;%
\begin{equation}
(i\overrightarrow{\partial_{1}}\cdot\gamma-m)\chi(x_{1}x_{2}\text{:}%
B)(i\overleftarrow{\partial_{2}}\cdot\gamma-m)=-e^{2}D_{F}^{\mu\nu}%
(x_{1}-x_{2})\gamma_{\mu}\chi(x_{1}x_{2}\text{:}B)\gamma_{\nu},
\end{equation}
where $m$ is a dynamical mass and $D_{F}$ is a photon propagator in the
covariant gauge%
\begin{equation}
D_{F}^{\mu\nu}(p)=\frac{g^{\mu\nu}}{p^{2}+i\epsilon}+(\xi-1)\frac{p_{\mu
}p_{\nu}}{p^{4}}.
\end{equation}
In momentum space we transform to center of mass and relative coordinate%
\begin{equation}
X=(x_{1}+x_{2})/2,x=x_{1}-x_{2},
\end{equation}%
\begin{equation}
P=p_{1}+p_{2},p=(p_{1}-p_{2})/2.
\end{equation}
Then we define $\chi(P,p)$,the Fourier transform of the Feynman amplitude,by%
\begin{equation}
\chi(x_{1}x_{2}\text{:}B)=\exp(iP\cdot X)\int d^{4}p\exp(ip\cdot x)\chi(P,p).
\end{equation}
The BS equation in momentum space assume the form%
\begin{equation}
((\frac{P}{2}+p)\cdot\gamma-m)\chi(P,p)(-(\frac{P}{2}-p)\cdot\gamma
-m)=-\frac{e^{2}}{(2\pi)^{4}}\int d^{4}qD_{F}^{\mu\nu}(p-q)\gamma_{\mu}%
\chi(P,p)\gamma_{\nu}%
\end{equation}
for fa system.Since under charge conjugation vector particle is odd,$\lambda$
should be replaced by $-\lambda.$For scalar case the $\chi^{S}(P,p)$ is
written in general which corresponds to spin singlet $S(P,p)$ and triplet
$V(P,p)$ and tensor $T(P,p)$
\begin{equation}
\chi^{S}(P,p)=S(P,p)I+P\cdot\gamma V^{1}(P,p)+p\cdot\gamma V^{2}%
(P,p)+\sigma_{\mu\nu}T(P,p)(P_{\mu}p_{\nu}-P_{\nu}p_{\mu}).
\end{equation}
For total momentum $P_{\mu}=0$ case the BS amplitdes are decoupled to
scalar-vector and tensor and we have a coupled equation for
scalar-vector$[2,6,7,8].$If we substitute eq (9) to eq (8) we get a
scalar-vector equation for Euclidean momentum $p$%
\begin{equation}
(m^{2}-p^{2})S(p^{2})-2p^{2}V^{2}(p^{2})=\lambda^{S}\int d^{4}q\frac{S(q^{2}%
)}{(p-q)^{2}},
\end{equation}
and
\begin{equation}
(m^{2}-p^{2})p_{\mu}V^{2}(p^{2})+2p_{\mu}S(p^{2})=\lambda^{V}\int d^{4}%
q\frac{q_{\mu}V^{2}(q^{2})}{(p-q)^{2}}.
\end{equation}%
\begin{align}
\lambda^{S}  &  =(3-\xi)\lambda...\text{ for ff-system}\nonumber\\
&  =-(3-\xi)\lambda..\text{ for fa-system},\\
\lambda^{V}  &  =2\xi\lambda..\text{for ff-system}\\
&  =-2\xi\lambda..\text{for fa-system,}%
\end{align}
where $\lambda=e^{2}/(4\pi)^{2}.$In the Landau gauge ($\xi=0$) $\lambda^{V}$
vanishes.The solution is given in ref[2].Then eq (11) reduces to a algebraic
equation and $V^{2}(p^{2})$ is obtained
\begin{equation}
V^{2}(p^{2})=\frac{-2}{(m^{2}-p^{2})}S(p^{2}).
\end{equation}
From eq (10) and (15) we obtain
\begin{equation}
\frac{(m^{2}+p^{2})^{2}}{m^{2}-p^{2}}S(p^{2})=\lambda^{S}\int d^{4}%
q\frac{S(q^{2})}{(p-q)^{2}}.
\end{equation}
By using the formula(Klein-Gordon equation for photon)%
\begin{equation}
\square_{p}\frac{1}{(p-q)^{2}}=-4\pi^{2}\delta^{(4)}(p-q)
\end{equation}
we get a differential equation for $S(p^{2})$ from eq (16)%
\begin{equation}
(s\frac{d^{2}}{ds^{2}}+2\frac{d}{ds})[\frac{(m^{2}+s)^{2}}{m^{2}%
-s}S(s)]=-\lambda^{S}S(s),
\end{equation}
where $s=p^{2}.$The solution is characterized by hypergeometric function%
\begin{align}
S(s)  &  =\frac{m^{2}-s}{(s+m^{2})}(s+m^{2})^{-1-\gamma/2}F(\alpha
,\beta,\gamma;\frac{m^{2}}{m^{2}+s}),\nonumber\\
\alpha &  =\frac{2+\sqrt{1+4\lambda^{S}}-\sqrt{1+8\lambda^{S}}}{2},\beta
=\frac{2+\sqrt{1+4\lambda^{S}}+\sqrt{1+8\lambda^{S}}}{2},\nonumber\\
\gamma &  =1+\sqrt{1+4\lambda^{S}},
\end{align}
which satisfy the boundary condition%
\begin{align}
\lim_{S\rightarrow\infty}(s\frac{d}{ds}+1)[\frac{(m^{2}+s)^{2}}{m^{2}-s}S(s)]
&  =0,\nonumber\\
\lim_{S\rightarrow0}s^{2}\frac{d}{ds}[\frac{(m^{2}+s)^{2}}{m^{2}-s}S(s)]  &
=0.
\end{align}
After angular integration eq (16) is rewritten%
\begin{equation}
\frac{(m^{2}+s)^{2}}{m^{2}-s}S(s)=\frac{\lambda_{S}}{2s}\int_{0}^{s}%
ds^{\prime}s^{\prime}S(s^{\prime})+\frac{\lambda_{S}}{2}\int_{s}^{\infty
}ds^{\prime}S(s^{\prime}).
\end{equation}
If we differentiate the above integral equation we get the boundary
conditions.In the case of $\alpha(\beta)=-n,F$ is a$\ n$-th degree of
hypergeometric series in $s.$We have a discrete set of spectrum if
$\lambda^{S}>0$(fermion-fermion system) and the eigenvalues are given%
\begin{equation}
\lambda_{n}^{S}=(n+1)[3(n+1)+\sqrt{8n^{2}+16n+9}],n=0,1,2...
\end{equation}
For the lowest eigenvalue $n=0$ $(\lambda_{0}=\lambda_{0}^{S}/3=2),$we have%
\begin{equation}
\chi_{0}(p)=\frac{m^{2}-p^{2}+2p\cdot\gamma}{(m^{2}+p^{2})^{5}}.
\end{equation}
For fermion-antifermion system there exists contineous spectrum for
$-1/8\leq\lambda<0$%
\begin{equation}
S(s)=\frac{m^{2}-s}{(s+m^{2})}(s+m^{2})^{-1-\gamma/2}F(\alpha,\beta
;\alpha+\beta-\gamma+1;\frac{s}{s+m^{2}}).
\end{equation}
Here we show the profile of $\chi_{0}(x)$ in FIG1.Fourier transformation of
$\chi(p)$ is defined%
\[
\chi(x)=\frac{1}{4\pi^{2}}\int p^{3}dp\frac{J_{1}(px)}{px}\chi(p),
\]%
\begin{align}
\chi_{0S}(x)  &  =\frac{x^{3}}{768\pi^{2}}K_{1}(x),\nonumber\\
\chi_{0V}(x)  &  =\frac{x}{1536\pi^{2}}(4xK_{0}(x)+(x^{2}+8)K_{1}(x)),
\end{align}
for $m=1.$%
\begin{figure}
[ptb]
\begin{center}
\includegraphics[
height=2.8283in,
width=2.8283in
]%
{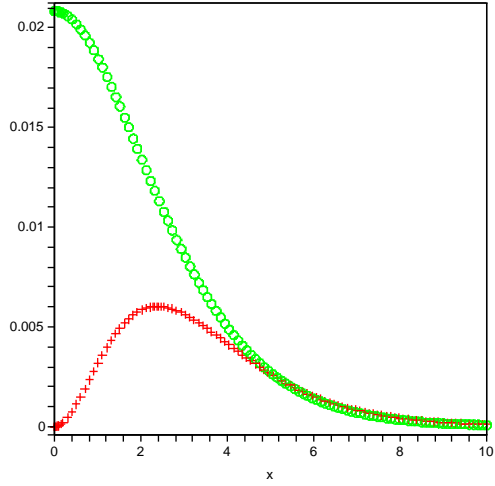}%
\caption{$4\pi^{2}\chi_{0}^{S}(x)(\times),4\pi^{2}\chi_{0}^{V}(x)(\bigcirc)$
for $m=1$}%
\label{f1}%
\end{center}
\end{figure}
In ref[4] introducing auxiliriary field for comosite operator effective action
has been made.If we minimize it we can derive the Dyson-Schwinger equation for
fermion and Bethe-Salpeter equation for boundstate.However for the first
approximation to the lowest excitation,vector was neglected.Thus the BS
equation for scalar meson is not a coupled equation with vector meson as eq(10),(11).

\subsection{\bigskip Massless boson in QED$_{3}$}

In ref[3] three dimensional QED is considered as an effective model which
discribes the two dimensional superconductor,where gap has nodes and the
low-energy fermionic excitations have linear dispersion.The fluctuating phase
of superonductor is considered as Berry gauge fields which couples to spin
degree of freedom of exciton.For these systems we assume the validity of
relativistic treatment for fermion.For definiteness we use four-dmensional
representation of $\gamma$ matrix [5].In ref[3] following type of equation for
the eigenvalue $e_{n}(\left\vert p\right\vert ,m)$ and normalized
eigenfunction $\psi_{n}(r,\left\vert p\right\vert ,m)$ for electron-hole
system near the nodes were dicsussed%
\begin{equation}
\lbrack-\partial\cdot\partial-V_{eff}]\psi_{n}=e_{n}\psi_{n},
\end{equation}%
\begin{equation}
\chi(p)=\frac{1}{16}\int d^{3}r\exp(-ip\cdot r)\sum_{ij}\left\langle
\overline{\psi}_{i}(r)\psi_{i}(r)\overline{\psi}_{j}(0)\psi_{j}%
(0)\right\rangle ,
\end{equation}%
\begin{equation}
\chi(p)=N\sum_{n}\frac{\left\vert \psi_{n}(0,\left\vert p\right\vert
,m)\right\vert ^{2}}{4e_{n}(\left\vert p\right\vert ,m)+p^{2}},
\end{equation}
where $V_{eff}$ is a potential by dressed gauge boson and has an invers-square
form for small $r$ and $\chi(p)$ is a sum of boundstate propagator with
residue $\left\vert \psi_{n}\right\vert ^{2}$of each pole$.$The conclusion is
that there are infinite numbers of negative energy.Here we simpliy notice the
results
\begin{align}
\lbrack-\partial\cdot\partial-\frac{\lambda}{16z^{2}}]\psi(z) &  =-\frac{1}%
{4}\psi,\lambda=(2+\xi)g^{2},\\
\psi(z) &  =\frac{K_{\sqrt{4-\lambda}/4}(z/2)}{\sqrt{z}}.
\end{align}
It is said that the continuum spectrum that is normalizable into a single
boundstate would be a so called conformal tower%
\begin{equation}
\psi_{0}(z)=\sqrt{\frac{\kappa^{3}}{2\pi}}\frac{K_{0}(\kappa\left\vert
z\right\vert )}{\sqrt{\kappa\left\vert z\right\vert }},
\end{equation}
where $-\kappa^{2}$ is the renormalized bound-state energy,and $K_{0}$ is the
modified Bessel function of the second kind.Apart from these realistic
application to condensed matter physics it is interesting to solve the
massless boundstate problems in three dimension.Now we return to relativistic
BS equation for fa system to study the same problems.For $P_{\mu}=0,T$ does
not couple to $S,V$.We consider the equation as (16) in the previous section.%
\begin{equation}
\frac{(m^{2}+p^{2})^{2}}{m^{2}-p^{2}}S(p^{2})=\lambda^{S}\int d^{3}%
q\frac{S(q^{2})}{(p-q)^{2}}.
\end{equation}
However we cannot solve the above equation as in four dimension since the
identity is%
\begin{equation}
\square_{p}\frac{1}{|p-q|}=-4\pi\delta^{(3)}(p-q)
\end{equation}
in this case.Therefore we expand the left hand side of the eq (32) in $p$ for
large and small cases.We get
\begin{align}
(-2m^{2}-p^{2})S(p) &  =\lambda^{S}\int d^{3}q\frac{S(q^{2})}{(p-q)^{2}%
},(m^{2}\ll p^{2}),\\
(m^{2}+2p^{2})S(p) &  =\lambda^{S}\int d^{3}q\frac{S(q^{2})}{(p-q)^{2}}%
,(p^{2}\ll m^{2}),
\end{align}
where $\lambda^{S}=e^{2}/2\pi.$ We convert these equations in position space%
\begin{align}
(-2m^{2}+\square_{x})S_{S}(x) &  =\lambda^{S}\frac{S(x)_{S}}{\left\vert
x\right\vert },(m\left\vert x\right\vert \ll1),\\
(m^{2}-2\square_{x})S_{L}(x) &  =\lambda^{S}\frac{S(x)_{L}}{\left\vert
x\right\vert },(1\ll m\left\vert x\right\vert ).
\end{align}
First we solve the following equation for scalar-vector ff system $S(x)$
\begin{align}
(m^{2}-\square_{x})S(x) &  =-\lambda^{S}\frac{S(x)}{\left\vert x\right\vert
},\lambda^{S}=\frac{e^{2}}{2\pi},\nonumber\\
\square_{x} &  =\frac{d^{2}}{dr^{2}}+\frac{2}{r}\frac{d}{dr}+\frac
{l(l+1)}{r^{2}}.
\end{align}
The relation among $S(x),S_{S}(x),S_{L}(x)$ may be clear by scaling the mass
or coupling constant.For simplicity we set $m=1.$For the ground state $l=0,$we
have
\begin{equation}
\frac{d^{2}S}{d^{2}r}+\frac{2}{r}\frac{d^{2}S(r)}{d^{2}r}-S(r)-\frac
{\lambda^{S}}{r}S(r)=0.
\end{equation}
For large $r$ if we neglect terms which are proportional to $1/r,$we obtain%
\begin{equation}
\frac{d^{2}S(r)}{d^{2}r}=S(r),
\end{equation}
from this $S(r)$ behaves as $e^{-r}.$For small $r$
\[
S(r)=u(r)/r.
\]
then the eq (39) becomes
\begin{equation}
\frac{d^{2}u}{dr^{2}}+(-1-\frac{\lambda}{r})u=0.
\end{equation}
Solutions of this equation must not diverge faster than the finite power of
$r$ and finite at $r=0.$The solution which satisfies this latter condition is
derived by Whittaker equation
\begin{equation}
\frac{d^{2}W}{d^{2}z}+(-\frac{1}{4}-\frac{\lambda}{z}+\frac{1/4-\mu^{2}}%
{z^{2}})W(z)=0,
\end{equation}
for $\mu=1/2,$whose solution is expressed by linear combination of
$M_{\lambda,1/2}$ and $W_{\lambda,1/2}$
\begin{equation}
S(x)=C_{1}\frac{M_{-\lambda/2,1/2}(2\left\vert x\right\vert )}{\left\vert
x\right\vert }+C_{2}\frac{W_{-\lambda/2,1/2}(2\left\vert x\right\vert
)}{\left\vert x\right\vert }.
\end{equation}
$S(x)$ is expanded near origin
\begin{equation}
S(x)_{\left\vert x\right\vert \rightarrow0}=\frac{C_{2}}{\left\vert
x\right\vert \Gamma(1+\lambda/2)}+2C_{1}+..
\end{equation}
Therefore for normalizability of $S(x)$ it is sufficent that $1/\Gamma
(1+\lambda/2)$ vanishes.This condition is realized for
\begin{equation}
1+\lambda/2=-n(n=0,1,2,..).
\end{equation}
This is an eigenvalue condition for the coupling constant $\lambda$.Thus
fermion-antifermion(fa) system has discrete spectrum($\lambda<0).$In the case
of positive copuling the function $M$ diverges as $\exp(x/2)[9].$Thus
fermion-antifermion(ff) system has no bound states($\lambda>0)$. Since the
coupling constant has a dimension of mass,the mass of the ground sate is
largest for fixed coupling $\lambda$.Here we return to $S_{S}(x)$ and
$S_{L}(x).$We have%
\begin{align}
S_{S}(x) &  =\frac{C_{1}M_{-a,1/2}(2\sqrt{2}m\left\vert x\right\vert
)}{\left\vert x\right\vert }+\frac{C_{2}W_{-a,1/2}(2\sqrt{2}m\left\vert
x\right\vert )}{\left\vert x\right\vert },\\
S_{L}(x) &  =\frac{D_{1}M_{a,1/2}(\sqrt{2}m\left\vert x\right\vert
)}{\left\vert x\right\vert }+\frac{D_{2}W_{a,1/2}(\sqrt{2}m\left\vert
x\right\vert )}{\left\vert x\right\vert },a=\frac{\sqrt{2}\lambda^{S}}{4m}.
\end{align}
Eigenvalues are given as
\[
\frac{\sqrt{2}\lambda^{S}}{4m}=-n(n=1,2....).
\]
For the large distance in the case of these negative $\lambda^{S},M_{a}%
,_{1/2}(\sqrt{2}m\left\vert x\right\vert )$ blows up.Thus we choose $D_{1}%
=0.$Here we have an approximate solution of the integral equation for short
and long distance%
\begin{align}
S(x) &  =\frac{C_{1}M_{-a,1/2}(2\sqrt{2}m\left\vert x\right\vert )}{\left\vert
x\right\vert }(m\left\vert x\right\vert \ll1)\\
&  =\frac{D_{2}W_{a,1/2}(\sqrt{2}m\left\vert x\right\vert )}{\left\vert
x\right\vert }(1\ll m\left\vert x\right\vert ).
\end{align}
We choose $C_{1}=1/2$ for normalization of $S(x).$For small $n$ we have an
explicit form of $M,W$ in terms of $z=\sqrt{2}m\left\vert x\right\vert $
\begin{align}
M_{1,1/2}(z)/z &  =\exp(-z)2,\nonumber\\
M_{2,1/2}(z)/z &  =\exp(-z)2(1-z),(z\ll1).
\end{align}%
\begin{align*}
W_{-1},_{1/2}(z)/z &  \simeq\exp(-z)/2z^{2},\\
W_{-2,1/2}(z)/z &  \simeq\exp(-z)/z^{3},(1\ll z).
\end{align*}
Since at long distance the function $W$ strongly dumps,we cut-off long
distance and the solution may be approximated
\begin{equation}
S(z)_{n}=M_{\lambda,1/2}(2\left\vert z\right\vert )/2\left\vert z\right\vert
,\lambda=n.
\end{equation}
If we transform into momentum space we get for $n=1,2,\sqrt{2}m=1$%
\begin{align}
\chi_{1}(p^{2}) &  =\int x^{2}d\left\vert x\right\vert \frac{\sin(p\left\vert
x\right\vert )}{(p\left\vert x\right\vert )}S(\left\vert x\right\vert
)_{1}=\frac{1}{(p^{2}+1)^{2}}(1-\frac{2\gamma\cdot p}{1-p^{2}}),\nonumber\\
\chi_{2}(p^{2}) &  =\frac{16\pi(-1+p^{2}+2p\cdot\gamma)}{(p^{2}+1)^{3}}.
\end{align}
Here we notice that the vector part of $\chi_{1}$ is not
normalizable,therefore we avoid $n=1$ case for scalar-vector fa systems.Thus
the ground state must be given by $n=2.$Correct solution in the whole region
may be evaluated by numerical analysis of integral equation (32) with angular
intergral
\begin{equation}
\frac{(m^{2}+p^{2})^{2}}{m^{2}-p^{2}}S(p)=\frac{\lambda}{(2\pi)^{2}p}\int
_{0}^{\infty}qdqS(q)\ln(\frac{p+q}{\left\vert p-q\right\vert }),\frac
{\lambda\sqrt{2}}{4m}=2,
\end{equation}
with an input of Fourier transform of eq (49)
\begin{equation}
S_{S}(p)=\frac{16(-2m^{2}+p^{2})}{(p^{2}+2m^{2})^{3}}%
\end{equation}
in the right hand side of eq (53).We see the profile of $\chi_{2}(x)$ in
FIG2.
\begin{figure}
[ptb]
\begin{center}
\includegraphics[
height=2.8283in,
width=2.8283in
]%
{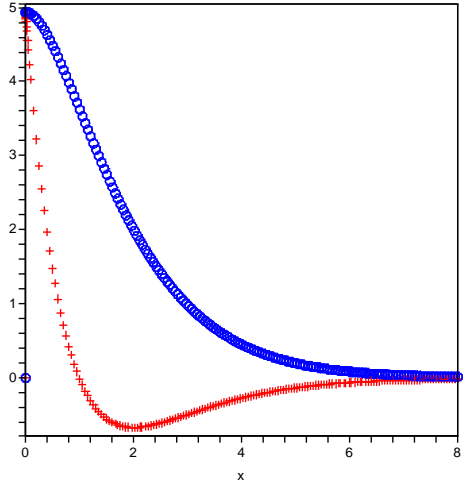}%
\caption{$\chi_{2}^{S}(x)(\times),\chi_{2}^{V}(x)(\square)$ for $m=1$}%
\label{f2}%
\end{center}
\end{figure}

\section{\bigskip Ward-Takahashi-identity for Axialvector currents}

In this section we examine the Ward-Takahashi-identity for the axialvector
currents.In our approximation to BS equation fermion mass is assumed to be
dynamical.Therefore if we have a solution of the BS equation for psedoscalar
it must satisfy the Ward-Takahashi-identity for the conservation of
axialvector currents[13]
\begin{equation}
\lim_{q\rightarrow0}q_{\mu}S_{F}(p^{\prime})\Gamma_{5\mu}(p^{\prime}%
,p)S_{F}(p)=\{S_{F}(p),\gamma_{5}\}\neq0,
\end{equation}
where $S_{F}(p)$ is a solution of \ the Dyson-Schwinger equation or
non-perturbative solution of the fermion propagator.The vertex function
$\Gamma_{5\mu}$ has a massless pole $q_{\mu}/q^{2}\chi^{P}(q),$where $\chi
^{P}(q)$ is a BS amplitude of psedoscalar Goldstone boson as a consequence of
chiral symmetry breaking. Following the notation in ref[5],$U(2)$ chiral
symmetry is generated by \{$I,\gamma_{4},\gamma_{5},\gamma_{45}\}$ which is
broken by dynamical fermion mass to a $U(1)\times U(1)$ symmetry generated by
$\{I,\gamma_{45}\}$ for the degree of freedom of scalar and psedoscalar as
$\{\sigma,\pi\}$ $.$The set of $\gamma$ matrices is%
\begin{align}
\gamma_{0}  &  =\left(
\begin{array}
[c]{cc}%
\sigma_{3} & 0\\
0 & -\sigma_{3}%
\end{array}
\right)  ,\gamma_{1,2}=\left(
\begin{array}
[c]{cc}%
\sigma_{1,2} & 0\\
0 & -\sigma_{1,2}%
\end{array}
\right)  ,\gamma_{4}=\left(
\begin{array}
[c]{cc}%
0 & I\\
-I & 0
\end{array}
\right)  ,\gamma_{5}=\left(
\begin{array}
[c]{cc}%
0 & -iI\\
iI & 0
\end{array}
\right)  ,\nonumber\\
\gamma_{45}  &  =-i\gamma_{4}\gamma_{5},\{\gamma_{\mu},\gamma_{4}%
\}=0,\{\gamma_{\mu},\gamma_{5}\}=0
\end{align}
Currents $\{\gamma_{\mu4},\gamma_{\mu5}\}$ are a analog of axialvector
currents and have a doublet of Goldstone boson which is called axial-scalar,%

\begin{equation}
\left(
\begin{array}
[c]{c}%
\chi^{(4)}(P,q)\\
\chi^{(5)}(P,q)
\end{array}
\right)  ^{AS}=\left(
\begin{array}
[c]{c}%
\gamma_{4}\\
\gamma_{5}%
\end{array}
\right)  S+\left(
\begin{array}
[c]{c}%
\gamma_{\mu4}\\
\gamma_{\mu5}%
\end{array}
\right)  (V^{1}P_{\mu}+V^{2}q_{\mu})+\epsilon_{\mu\nu\rho}P^{\mu}q^{\nu
}\left(
\begin{array}
[c]{c}%
\gamma^{\rho5}\\
-\gamma^{\rho4}%
\end{array}
\right)  T.
\end{equation}
Scalar-PS($\gamma_{45})$ BS amplitude is%
\begin{equation}
\chi^{S}(P,q)=S+\gamma\cdot qV^{2}+\gamma\cdot PV^{1}+\epsilon_{\mu\nu\rho
}P^{\mu}q^{\nu}\gamma^{\rho45}T.
\end{equation}%
\begin{equation}
\chi^{PS}(P,q)=S(P,q)\gamma_{45}+q^{\mu}\gamma_{\mu45}V^{2}+P^{\mu}\gamma
_{\mu45}V^{1}+\epsilon_{\mu\nu\rho}P^{\mu}q^{\nu}\gamma^{\rho}T.
\end{equation}
To check the Ward-Takahashi-identity we consider the equation for axial-scalar%
\begin{align}
(m^{2}+p^{2})\chi^{AS}(p)  &  =\frac{\lambda^{AS}}{(2\pi)^{3}}\int
d^{3}p^{\prime}\frac{\chi^{AS}(p^{\prime})}{(p-p^{\prime})^{2}},\\
(m^{2}-\square_{x})\chi^{AS}(x)  &  =\lambda^{AS}\frac{\chi^{AS}%
(x)}{\left\vert x\right\vert },\lambda^{AS}=\lambda.
\end{align}
for ff system.Sign of the coupling is opposite to the one in even
dimension.This is the same equation discussed in scalar-vector case and we
have solutions
\begin{align}
\chi^{AS}(x)  &  =\frac{M_{\lambda/2m,1/2}(2m\left\vert x\right\vert )}%
{|x|},\nonumber\\
\chi_{1}^{AS}(p)  &  =\frac{1}{(p^{2}+m^{2})^{2}}.
\end{align}
for $\lambda/2m=n(1,2...)$.Following the Ward-Takahashi-identity $\chi
^{AS}(x)$ for fa is derived from the scalar part of the fermion
propagator.Dyson-Schwinger equation for fermion propagator is written%
\begin{align}
\Sigma(p)  &  =\frac{e^{2}}{(2\pi)^{3}}\int d^{3}k\gamma_{\mu}\frac
{k\cdot\gamma+M(k)}{k^{2}+m^{2}}\gamma_{\nu}\frac{g_{\mu\nu}-(p-k)_{\mu
}(p-k)_{\nu}/(p-k)^{2}}{(p-k)^{2}},\\
M(p)  &  =\frac{e^{2}}{4\pi^{2}}\int d^{3}k\frac{M(k)}{k^{2}+m^{2}}\frac
{1}{(p-k)^{2}}%
\end{align}
in the Landau gauge with linear approximation $m=M(0)$.We define $F(x)$ as the
Fourier transformation of $M(p)/(p^{2}+m^{2})$
\begin{align}
F(x)  &  =\int\frac{d^{3}p}{(2\pi)^{3}}\exp(ip\cdot x)\frac{M(p)}{p^{2}+m^{2}%
},\\
(-\square_{x}+m^{2})F(x)  &  =\frac{2e^{2}}{(2\pi)^{6}}\int d^{3}pe^{ip\cdot
x}\int d^{3}p^{\prime}\frac{M(p^{\prime})}{p^{\prime2}+m^{2}}\frac{1}{k^{2}%
}\nonumber\\
&  =2e^{2}\int\frac{d^{3}p^{\prime}}{(2\pi)^{3}}e^{ip^{\prime}\cdot x}%
\frac{M(p^{\prime})}{p^{\prime2}+m^{2}}\int\frac{d^{3}k}{(2\pi)^{3}}%
\frac{e^{ik\cdot x}}{k^{2}}\nonumber\\
&  =2e^{2}\frac{1}{4\pi\left\vert x\right\vert }F(x),
\end{align}
where $k=p^{\prime}-p$ and used that $\int d^{3}p=\int d^{3}k.$We arrive at a
Schr\"{o}dinger like equation for $F(x)$
\begin{equation}
(-\square_{x}+m^{2})F(x)=\lambda\frac{F(x)}{\left\vert x\right\vert }%
,\lambda=\frac{e^{2}}{2\pi}.
\end{equation}
Its solution is given as the groudstate of hydrogen atom
\begin{equation}
F(x)=\frac{m^{2}}{8\pi}\exp(-m\left\vert x\right\vert ),F(p)=\frac{m^{3}%
}{(p^{2}+m^{2})^{2}}.
\end{equation}
In this case the ground state solution corresponds to $\lambda/2m=1(m=e^{2}%
/4\pi),$where mass is largest for fixed coupling[10].We also meet this
condition in the analysis of the fermion propagator based on low-energy
theorem[11,12].In fact we can determine a precise infrared behaviour of the
propagator in the above analysis but do not have a definite ultraviolet
behaviour.However if we demand the nonvanishment of the order parameter
$\left\langle \overline{\psi}\psi\right\rangle \neq0$ ,we obtain the condition
for the anomalous dimension of the wave function which must be unity.This
condition is approximately satisfied in the case of $1/N$ correction for
photon propagator with vanishing bare mass in the Landau gauge[12].This is
just the statement of Nambu-Goldstone theorem[14,15].

\section{Summary}

In this work we studied the solutions of spinor-spinor Bethe-Salpeter equation
for massless boson in QED$_{3}$ with ladder approximation in the Landau
gauge.It is not easy to solve this equation directly in three
dimension.However we use the approximate form of the integral equation for
large and small momentum respectively.These are converted to the differential
equation which are similar to the Schr\"{o}dinger type equation.In this case
an eigenvalue are determined by boundary condition at $x=0.$Thus we obtain
solutions for massless scalar-vector fa systems with discrete spectrum.In our
approximation massless scalar-vector ff system does not seem to exist in three
dimension.Finally it is shown that Ward-Takahashi-identity for axial currents
is satisfied with the Dyson-Schwinger equation for the fermion propagator in
the linear approximation to dynamical mass .

\section{References}

[1]J.R.Schrieffer,\textbf{Theory of Superconductivity};Westview press(1999).

[2]K.Higashijima,Prog.Theor.Phys.\textbf{55}(1976)1591;

\ \ \ A.Nishimura,K.Higashijima,Prog.Theor.Phys.\textbf{56}(1976)908.

[3]Babak H.Seradjeh,Igor F.Herbut:EprintArXiv:cond-mat/0701724;

\ \ \ I.F.Herbut,Physical Review Letters \textbf{94,}%
{\normalsize 237001(2005).}

[4]T.Morozumi,H.So,Prog.Theor.Phys.\textbf{77}(1987)1434.

[5]W.E.Allen,C.J.Burden,Phys.Rev.\textbf{D53},5482(1996).

[6]K.Nishijima,\textbf{Fields and Particles};W.A.BENJAMIN.INC(1969).

[7]N.Nakanishi,Prog.Theor.Phys.Pphys.Supplement,\textbf{43}(1969).

[8]V A Miransky,DYNAMICAL SYMMETRY\ BREAKING IN QUANTUM FIELD

\ \ \ THEORIES,World Scientific(1993).

[9]M.Abramowitz,I.A.Stegun,HANDBOOK OF MATHEMATEICAL FUNCTIONS;

\ \ \ DOVER\ PUBLICATIONS,INC.(1972).

[10]M.Koopmans,DYNAMICAL MASS GENERATION IN QED$_{3},$

\ \ \ \ Ph.D thesis in Groningen University(1990),

\ \ \ \ K.Higashijima,Prog.Theor.Phys.Supplement,\textbf{104(}%
{\normalsize 1991)}.

[11]Y.Hoshino,\textbf{JHEP}0409:048,2004.

[12]Y.Hoshino,EprintArXive:hep-th/0610016.to be published in
Nucl.Phys\textbf{.A.}

[13]T.Maskawa,H.Nakajima,Prog.Theor.Phys.\textbf{52}(1974)1326.

[14]J.Goldstone,Nuov.Cim.\textbf{19}(1961)154.

[15]Y.Nambu and G.Jona-lasinio,Phys.Rev.\textbf{122}(1961)345.
\end{document}